\documentclass[12pt,preprint]{aastex}

\def \etal {et al.~}
\def \apj {ApJ}

\def \solphys {Solar Phys.}

\def \aap {A\&A}

\begin{document}

\title{Simulation of quiet Sun waves in the Ca II infrared triplet}

\author{A. Pietarila\altaffilmark{1}, H. Socas-Navarro and T. Bogdan}
\affil{High Altitude Observatory, National Center for Atmospheric Research\altaffilmark{2}, 3080 Center Green, Boulder, CO 80301, USA\vbox{}}
 \author{M. Carlsson}
       \affil{Institute of Theoretical Astrophysics, University of Oslo, P.O.Box 1029 Blindern, N-0315 Oslo, Norway\vbox{}}
\author{R.F. Stein}
       \affil{Department of Physics and Astronomy, Michigan State University, East Lansing, MI 48824, USA\vbox{}}
       
\altaffiltext{1}{Institute of Theoretical Astrophysics, University of Oslo, P.O.Box 1029 Blindern, N-0315 Oslo, Norway}
\altaffiltext{2}{The National Center for Atmospheric Research (NCAR) is sponsored by the National Science Foundation.}

\begin{abstract}
%
The Ca II infrared triplet around 8540 \AA\ is a good candidate for
observing chromospheric magnetism. We show results from combining a
radiation hydrodynamic simulation with a Stokes synthesis code. The
simulation shows interesting time-varying behavior of the Stokes $V$
profiles as waves propagate through the lines' formation
regions. Disappearing and reappearing lobes in the Stokes $V$ profiles
as well as profile asymmetries are closely related to the atmospheric
velocity gradients.
\end{abstract}
   

\section{Introduction}
\label{intro}

New observational data from the SOHO and TRACE spacecraft have revealed
a wide array of wave activity in the corona, for example, the
discoveries of EIT waves \citep{Thompson+others1998}, longitudinal
standing waves within coronal loops \citep{Kliem+others2002,Wang+others2002}, and compressible waves in polar plumes
\citep {Ofman+others1997, DeForest+Gurman1998, Ofman+others1999}. \citet{Aschwanden2004} and
\citet{Nakariakov2003} give extensive reviews of coronal oscillation
observations. These observations underscore the link between
wave propagation and magnetic fields.

Similar observational studies of the connection between wave processes
and magnetic fields in the chromosphere are challenging: there are few
suitable spectral lines and they form under complex non-LTE (nLTE)
conditions. Consequently there is no straightforward way to extract
information, such as atmospheric temperature and pressure, from these
observations. The presence of emission self-reversals in the Ca line
cores, due to upward-propagating shock-heated plasma, complicates even
the simplest Doppler shift analysis and obscures the true oscillatory
signal. The magnetic flux in the quiet Sun is very small ($\approx$ 3
Mx cm$^{-2}$ in the internetwork, \citet{Lites2002}), so useful
observations need a high signal-to-noise ratio (S/N $\geq$ $10^3$),
which, when combined with the high time cadence required to resolve
the wave propagation, imposes stringent demands on the
instrumentation.

Acoustic waves generated by the turbulent solar convection become
non-linear and develop into shocks as they propagate into the low
density of the chromosphere
\citep{Carlsson+Stein1997}. Understanding this complex process is
important not only from a fundamental point of view, but also because
it has been traditionally thought to be an efficient way to transport
energy into the upper atmosphere. In 1948 both Biermann and
Schwarzschild proposed that shock waves heat the chromosphere, but in
recent times, this heating mechanism has been challenged. Carlsson \&
Stein (1995) have argued, based on radiation hydrodynamic simulations,
as to whether acoustic shocks are capable of heating the non-magnetic
chromosphere. And recently Fossum \& Carlsson (2005) showed, by
comparing radiation hydrodynamic simulations and TRACE observations,
that the integrated acoustic wave power, including frequencies up to
50 mHz, is insufficient to heat the non-magnetic chromosphere.

Even if wave propagation is not the sole answer to the heating problem
it is still of interest. Atmospheric dynamics depend on wave
propagation properties such as wave generation, refraction,
reflection, mode coupling and dissipation. Because the three MHD
waves, i.e. Alfv\'en, slow MHD and fast MHD waves, have quite
different properties, a lot of information about the atmospheric
conditions are embedded in them. In the mid-chromosphere magnetic
fields spread out and fill the space, making the term non-magnetic
chromosphere applicable at best only in the low quiet Sun
chromosphere. The layer where the transition from pressure to
magnetic dominated regimes occurs is often referred to as the
``magnetic canopy.'' Study of TRACE observations has shown that there
is more oscillatory power below the canopy than above it
\citep{McIntosh+others2003} and MHD simulations of wave propagation
have identified this layer as a place where dramatic changes in wave
propagation take place \citep{Bogdan+others2003}. Waves are also
used as a tool in ``coronal seismology'' to probe the physical
conditions in the corona. Coronal parameters, e.g. magnetic field
strengths and transport coefficients, can be determined by combining
measurements of the wave properties and oscillations with theoretical
models.

Modeling wave propagation in the solar atmosphere is a daunting
task. A realistic model must be three-dimensional (3D). The treatment
of radiation should be non-gray and nLTE. The significance of magnetic
fields in wave propagation make them an essential part of any
comprehensive model. Yet another crucial part of any model describing
wave propagation in the chromosphere is nonlinear effects such as wave
steepening and shocks. Because of the complexity of the problem, and
the computational power needed to solve it, so far no fully
comprehensive model has been developed. Instead restricted
combinations of the above-mentioned elements have been studied. For
example Carlsson \& Stein's (1992, 1994, 1995, 1997, 2002) 1D model
combines the nonlinear hydrodynamics with full treatment of radiative
transfer. The effect of shock dissipation in the chromosphere has been
studied also by e.g. \citet{Fawzy+others2002},
\citet{Rammacher+Cuntz2003}, \citet{Buchholz+others1998}, and
\citet{Skartlien+others2000}. \citet{Bogdan+others2003} have looked at
the nonlinear non-radiative 2D magnetic problem. Other numerical
simulations on MHD wave propagation include e.g.:
\citet{Ziegler+Ulmschneider1997}, \citet{Cargill+others1997},
\citet{Tirry+Berghmans1997}, \citet{Huang+others1999},
\citet{Mckenzie+Axford2000}, \citet{DeGroof+Goossens2000},
\citet{DeGroof+others2002}, \citet{Tsiklauri+others2003},
\citet{Arregui+others2003}, and \citet{Ofman+others2000}. In this paper
we approach the problem from a different perspective.


We have combined full treatment of radiative transfer in 1D with
magnetohydrodynamics in 2D. The treatment of the interplay between the
hydrodynamics and the magnetic field is effected in a suggestive, although
not entirely rigorous fashion. A 1D radiation
hydrodynamic simulation (duration of 3690 s) is employed as the base building block. It is used
to kinematically deform a 2D initial magnetic field,
$\mathbf{B}_0 \textrm{(x,z)}$, according to the magnetic induction equation,

\begin{eqnarray}
\frac{\partial \mathbf{B}}{\partial t} = \nabla \times [ \mathbf{v}\textrm{(z,t)} \times \mathbf{B}\textrm{(x,z,t)}],
\label{eqn:induction}
\end{eqnarray}
where $\mathbf{B}\textrm{(x,z,t=0)} = \mathbf{B}_0\textrm{(x,z)}$ is
prescribed and $\mathbf{v}\textrm{(z,t)} = w\textrm{(z,t)}
\mathbf{e}_z$ is derived from the radiation hydrodynamic
simulation. By having the plasma-$\beta$, i.e. the ratio between the
gas and magnetic pressures, significantly larger than one, we can
justify neglect of the back reaction of the Lorentz force on the
dynamics. Finally we integrate the Stokes transfer equation to
determine the emergent Stokes profiles in I and V for the Ca II
infrared (IR) triplet lines for each 10 second time step. In this way,
all 3 essential ingredients are accounted for in this approach
although not fully consistently.

The \ion{Ca}{2} infrared triplet is an excellent diagnostic of
chromospheric oscillations in sunspots
\citep[e.g.][]{Tziotziou+others2002,Socas-Navarro+others2001,Beckers+Schultz1972}) and shows promise to be as useful in the
quiet Sun. Unfortunately, these lines are usually out of the
accessible wavelength range for most modern spectro-polarimetric
instrumentation. Currently the newly commissioned Spectro-Polarimeter
for INfrared and Optical Regions
\citep[SPINOR,][]{Socas-Navarro+others2005} and the French-Italian solar
telescope THEMIS operated in the MTR-mode \citep{Mein+Rayrole1985,Rayrole+Mein1993} can routinely provide high-sensitivity
observations of these lines.

The outline of the paper is as follows. In the second section the
numerical simulation and initial conditions are described. Then in the
third section we show the Stokes profiles produced in the simulation
and try to explain how they are formed in terms of the atmospheric
dynamics. We find interesting time-dependent behavior in the Stokes V
profiles as shocks propagate through the model
atmosphere. Disappearing and reappearing Stokes V lobes as well as
profile asymmetries are closely related to both the local fluid
velocity and atmospheric velocity gradients. Finally, in the fourth
section the results are discussed.

\section{The numerical simulation}
\label{outline}

Since solving the full 3D nonlinear problem is out of our reach we
chose a simplified approach with the main premise being that the
magnetic field is weak. This way we could construct a 2D
dynamic magnetized atmosphere without needing to take into account the
full coupling of the hydrodynamics and the magnetic field. Instead the
time evolution of the magnetic field is dictated entirely by the
hydrodynamics. 

In expanding a 1D hydrodynamic simulation into 2D
we made three assumptions. The first one is that the piston used to
drive the atmosphere is homogeneous over a larger area. In MDI
observations the size of coherently oscillating patches is typically
up to 6 Mm in diameter \citep{Judge+others2001}. Secondly, that
there is no horizontal radiative transfer or interference between
points. SUMER observations of the UV continua show a sharp contrast
between the network and internetwork. This suggests that the length
scale of lateral radiative transfer is not large i.e. it is smaller
than the SUMER spatial resolution of 1 arc second. Previous
simulations by Carlsson and Stein using the radiation hydrodynamic
code successfully reproduce observations of the low chromosphere, up
to about 1.2 Mm. This implies that the code, even though 1D, captures
most essential physical ingredients of the lower solar atmosphere. And
the third assumption is that the filling factor for the magnetic field
is 1, i.e. we are fully resolving it.

In the following subsections the three building blocks of the
simulation, i.e. the radiation hydrodynamic simulation, adding the
magnetohydrodynamics and synthesizing Stokes profiles, will be discussed
separately.

\subsection{Radiation hydrodynamic simulation}

We used Carlsson and Stein's (1992, 1995, 1997, 2002) radiation
hydrodynamic code as the starting point for the numerical
simulation. Not only did Carlsson and Stein succeed in reproducing the
appearance of the Ca II H and K lines, but they also showed beyond
doubt the dynamic nature of the solar atmosphere. Their code, which is
1D, non-magnetic, and driven by a piston derived empirically from
suitable time series observations, compares with observations of the
Ca II H and K bright grains remarkably well, even down to the level of
individual grains. They also showed that statistical equilibrium is
not a valid approximation for the calculation of hydrogen populations
in a dynamic atmosphere. Furthermore, they demonstrated that there is
fundamental difference between the static, semi-empirical
chromospheric models and a truly dynamic chromosphere.

The radiation hydrodynamic code solves simultaneously the mass,
momentum and energy conservation equations and the nLTE radiative
transfer equations on an adaptive grid. An adaptive grid allows for
higher numerical resolution in regions where it is needed,
i.e. shocks, and a coarser discretization in places where a lower
resolution is sufficient. Not all species contribute
significantly to the thermodynamics, only H, He, and Ca are included in
the full-scale radiation hydrodynamic treatment. All continua besides
H, Ca, and He are treated as background continua in LTE using the
Uppsala opacities code \citep{Gustafsson1973}.

\begin{figure*}
\epsscale{1}
\plotone{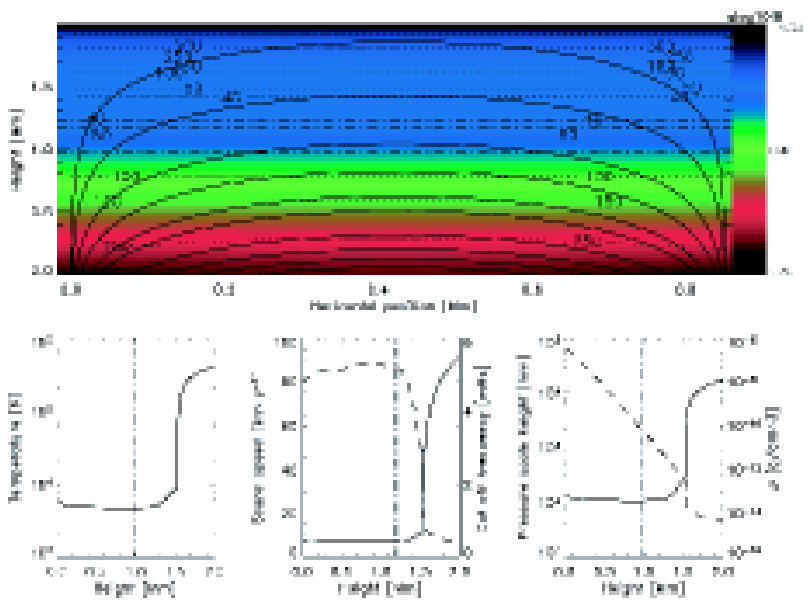}
\caption{The initial atmosphere. Upper panel shows the initial magnetic field configuration. The color scale indicates field strength on a logarithmic scale. Solid lines are representative magnetic field lines. Dotted lines are plasma-$\beta$ contours. Dash-dotted lines are the heights where optical depths in the Ca line cores equal unity. The lower panels show the temperature (left panel), sound speed and acoustic cut off frequency (center panel, the left ordinate is for the solid line and right ordinate for the dashed line), pressure scale height and density (right panel) as functions of height.   
\label{fig:atmos0}
}
\end{figure*}

A piston (fluid velocity as a function of time) based on MDI
observations is located at the lower boundary and is used to force the
atmosphere. Initially the atmosphere (figure \ref{fig:atmos0}) is in
radiative equilibrium. The radiation field at the lower boundary is
optically thick at all wavelengths while the upper boundary is
transmitting and is set to have a $10^{6}$ K temperature at a fixed
height of 10 Mm above the photosphere. In our simulation we used only the first 3.5 Mm since above that the Ca II IR triplet line opacities are small and the region does not contribute to line formation. The coronal irradiation of the
chromosphere is based on Tobiska's formalism
\citep{Wahlstrom+Carlsson1994}. For a more detailed discussion of
the lower boundary, see \citet{Judge+Carlsson+Stein2003}.

\begin{figure*}
\epsscale{1}
\plotone{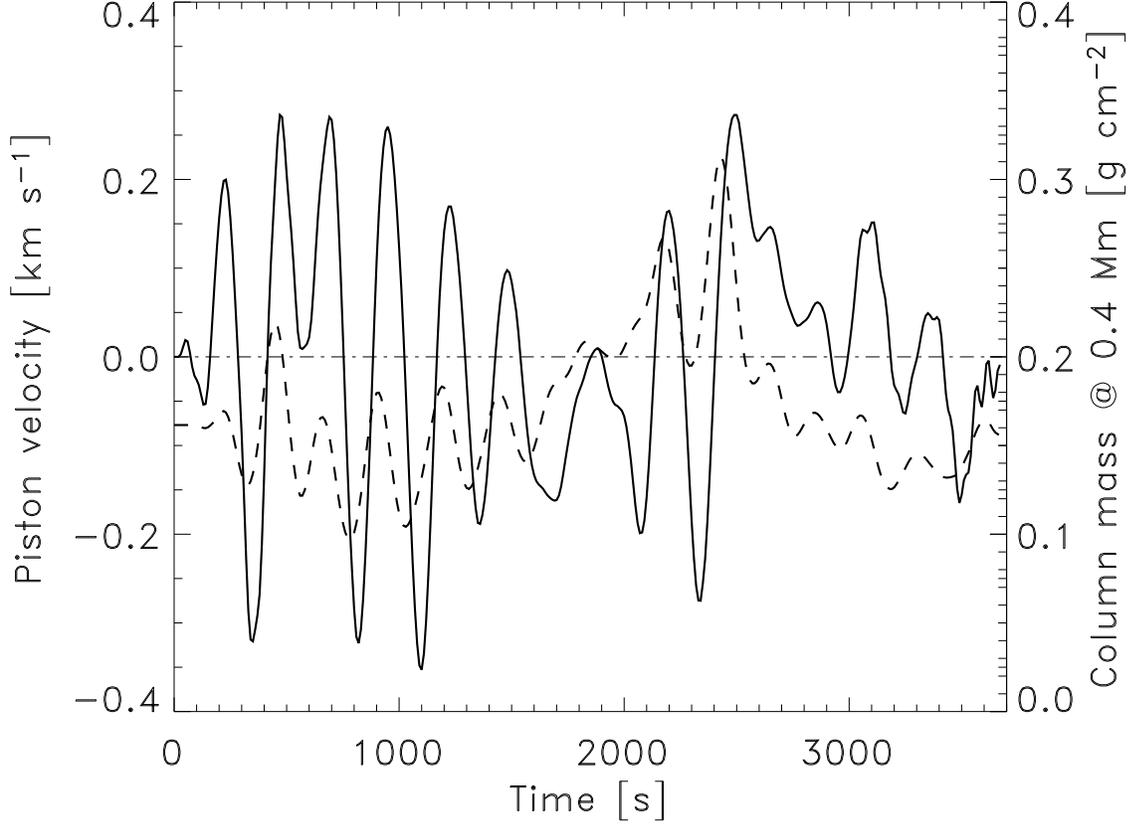}
\caption{The piston velocity at the lower boundary as a function of time. Negative and positive velocities indicate upward and downward motion, respectively. Note the asymmetry of the piston in the first half of the simulation with positive velocities dominating. The column mass at a fixed height, 0.4 Mm is also shown (dashed line, left ordinate).  
\label{fig:piston}
}
\end{figure*}

Figure \ref{fig:piston} shows the piston velocity used in
the simulation as a function of time. Negative velocities correspond to an 
upward movement of the
atmosphere. The main frequencies of the driving piston are centered around 3
mHz, i.e. frequency range for p-mode oscillations, which usually dominate the
MDI velocity power spectra. There is a slight temporal asymmetry in the piston,
with positive velocities being dominant in the first half of the time series. This
causes a slow downward movement of the whole atmosphere during the first
half of the simulation. This effect can be seen as a slow drift of the
column mass at a given height as a function of time (figure
\ref{fig:piston}).
 
\begin{figure*}
\epsscale{1}
\plotone{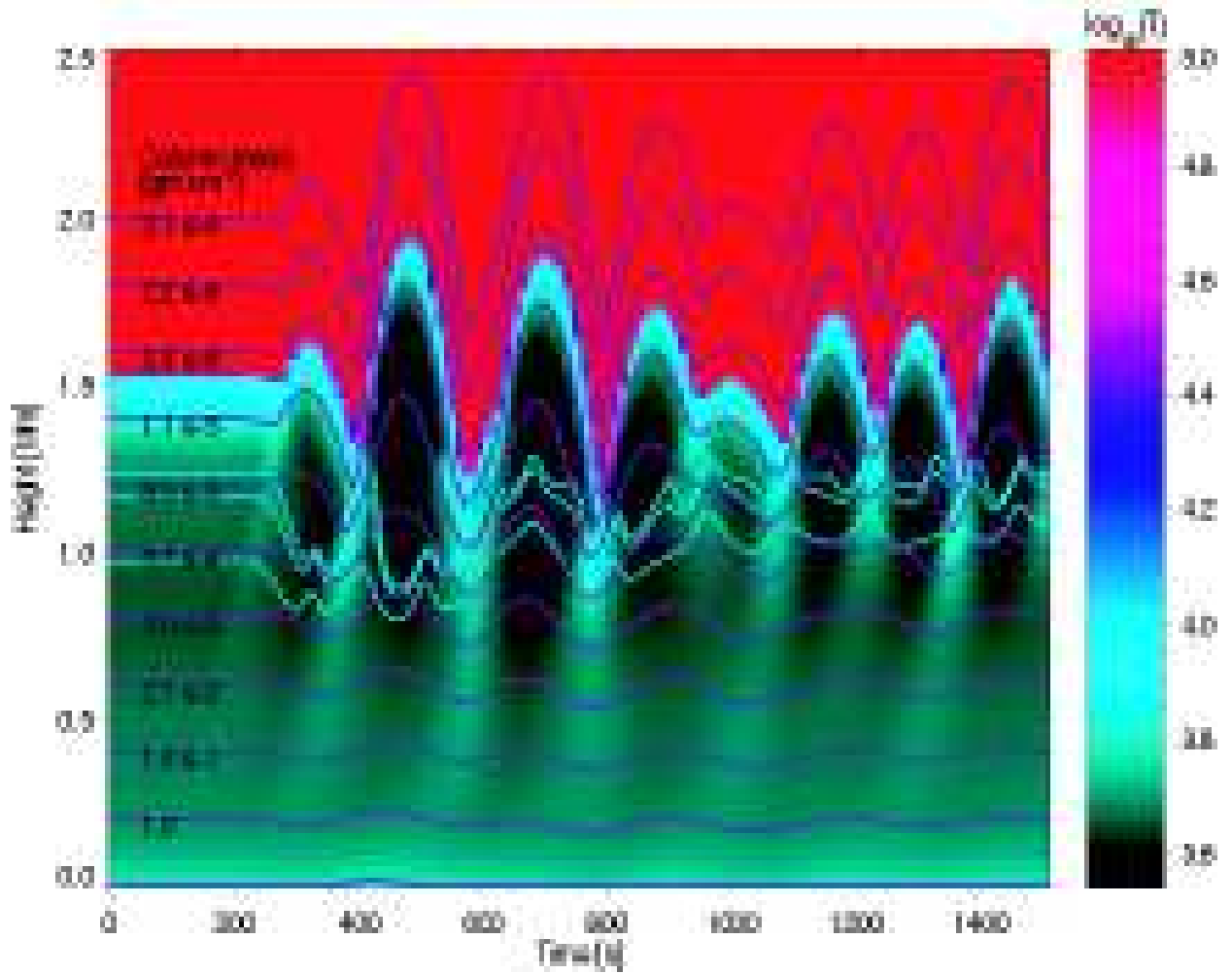}
\caption{The dynamic evolution of the first 1500 seconds in the radiation hydrodynamic simulation. The color scale of the image shows the temperature as a function of height (y-axis) and time (x-axis). Over-plotted in turquoise are the formation heights ($\tau_{\nu}=1$ at line core) of the Ca IR triplet lines (lowest line: 8498 \AA, middle:8662 \AA, top:8542 \AA) and in blue the locations of fixed column masses. The steepening of the waves into shocks as they propagate upward in the atmosphere is clearly seen in all three.
\label{fig:radyn-shock}
} 
\end{figure*}

As the density in the atmosphere decreases with height, the velocity
amplitudes of waves must increase as they propagate upward in order to
approximately conserve the wave energy flux. Eventually, by the
mid-chromosphere, the waves steepen and form shocks (figure
\ref{fig:radyn-shock}). As the large amplitude waves propagate they
push the less dense material in front of them upward compressing and
thus heating it. After the shock has passed the material cools and
falls down again. This motion is clearly seen in figure
\ref{fig:radyn-shock} where the location of a fixed column mass
oscillates as the waves pass by. This oscillatory behavior is
manifested dramatically by the periodical change in the location of
the large atmospheric temperature gradient, i.e. the transition
region. This variation can be up to 0.5 Mm in extent. Most of the heat
dissipated from the shock goes into thermal energy instead of
ionization because the dynamic timescales are much shorter than the
hydrogen ionization/recombination timescales. Hydrogen is by far the
most abundant element and therefore very important for the
thermodynamics. Ionization does not take place until sometime after
the shock front has passed. If LTE was assumed the
ionization/recombination would be instantaneous leading to a smaller
temperature variation in the shock as shown by
\citet{Carlsson+Stein1992}.

\subsection{Magnetohydrodynamical part of the numerical simulation}

The time evolution of the magnetic field is governed by the induction equation. Expanding the crossproduct in equation (\ref{eqn:induction})
gives:

\begin{eqnarray}
\frac {\partial \mathbf{B}}{\partial t}+ \left(\mathbf{v} \cdot \nabla \right )\mathbf{B} + \mathbf{B} \left( \nabla \cdot \mathbf{v} \right )= \left ( \mathbf{B} \cdot \nabla \right )\mathbf{v},
\label{eqn:induction2}
\end{eqnarray}
where $\mathbf{v}=w\textrm{(z,t)}\hat{\mathbf{e}}_{z}$. The $z$-component of equation (\ref{eqn:induction2}) in a 2D
vertically stratified atmosphere is:

\begin{eqnarray}
\frac{\partial B_z}{\partial t}+w \frac{\partial B_z}{\partial z} =\frac {dB_z}{dt}=0
\label{eqn:bz}
\end{eqnarray}
i.e. the Lagrangian derivative vanishes, but the Eulerian does not. The $x$-component of equation
(\ref{eqn:induction2}) is:

\begin{eqnarray}
\frac {\partial B_x}{\partial t} + v_z \frac{\partial B_x}{\partial z}+B_x \frac{v_z}{\partial z}=0.
\end{eqnarray}
Dividing by $\rho$, and then from this subtracting the equation for
conservations of mass,

\begin{eqnarray}
\frac{\partial \rho}{\partial t}+ w\frac{\partial \rho}{\partial z}+\rho \frac{\partial w}{\partial z} \cdot \mathbf{v}=0
\end{eqnarray}
multiplied by $B_x/\rho^2$ gives:

\begin{eqnarray}
\frac{d}{dt}\frac{B_x}{\rho}=0
\label{eqn:bx}
\end{eqnarray}
i.e. it is the Lagrangian derivative of $B_x/\rho$ that vanishes. The $B_x$ carried by a given fluid element evolves in the same manner as the density:
 
\begin{eqnarray}
\frac{B_{x}(z',0)}{B_{x}(z,t)}=\frac{\rho(z',0)}{\rho(z,t)},
\end{eqnarray}
where $z'$ is the location of the fluid parcel at $t=0$ and $z$ is the
position of that same fluid parcel at a later time $t$. The density
and velocity fields are furnished by the radiation hydrodynamic
simulation. With this information, and equations (\ref{eqn:bz}) and
(\ref{eqn:bx}) we can solve for the time evolution of the magnetic
field. By doing this we have a 2D magnetized dynamic atmosphere where
the hydrodynamics vary only in the vertical direction.


The choice of initial magnetic field is arbitrary as long as the main
criterion is fulfilled: the field has to be sufficiently weak. We chose
to start out with a potential field. It was constructed from the
first 20 coefficients of the Fourier series of a step function and
set to decay exponentially with height:

\begin{equation}
  \begin{array}{c}
    B_{x}\textrm{(x,z,t=0)} \\
    B_{z}\textrm{(x,z,t=0)}\\
  \end{array} 
  \left \} = \frac{2B_{0}}{\pi} \right. \sum_{k=0}^{20} \frac{e^{\frac{-\pi(2k+1)}{H}z}}{2k+1} \\ 
\left \{ \nonumber
\label{eqn:field}
    \begin{array}{c}
      \sin \left [\frac{\pi(2k+1)}{H}\times x \right ] \\
      \cos \left [\frac{\pi(2k+1)}{H}\times x \right ] \\
    \end{array}
    \right.,
\end{equation}

where $B_0$=150 G and $H$=0.85 Mm. The simulation
domain spans 0.85 Mm in the horizontal direction with 23 equidistant grid
points and about 3.5 Mm in the vertical direction with 130 adaptive
grid points. Note that 3.5 Mm is well above the formation height of the Ca
lines. 

The magnetic field configuration is quite simple (figure
\ref{fig:atmos0}). It contains one arcade-like feature that extends
from one edge of the domain to the other edge. Near the lower boundary
the field strength is 100-250 G, and is largest at the sides of the
domain where the field is nearly vertical. In the middle the field is
horizontal. The field decays rapidly with height: by 1 Mm the field
strength has reduced by roughly two orders of magnitude. The
variation of the magnetic pressure in the horizontal direction is by
then smoothed out, because the $k=0$ term dominates in equation
(\ref{eqn:field}) at great heights. Because the initial field is
potential, the magnetic field scale height is set by the width of the
magnetic loop structure, 0.85 Mm. The plasma-$\beta$ decreases with
height in the lower part of the simulation domain (the smallest value
in the initial atmosphere is roughly 40), but above 1.2 Mm it begins
to increase again. In the lower part, the gas pressure scale height is
smaller than the magnetic field scale height. So, the gas pressure
drops off faster than the field strength resulting in a decreasing
plasma-$\beta$. However, at about 1.2 Mm the magnetic field scale
height becomes smaller than the density scale height, which, unlike
the magnetic field scale height, is not constant throughout the
atmosphere. This causes the plasma-$\beta$ to increase with
height.

As the waves propagate through the atmosphere the magnetic field is
rarefied and compressed, and no longer potential. The smallest value
of plasma-$\beta$, 3, occurs at $\approx$ 800 s while the atmosphere
is still adjusting to the initial conditions. After this the
plasma-$\beta$ is in general over 10 except when large shocks pass by
and it decreases to 5-10. Keeping the plasma-$\beta$ large justifies
the neglect of the Lorentz force.

\subsection{Synthesizing the Ca IR triplet Stokes profiles}

The nLTE radiative transfer equation is both non-linear (the atomic
populations through the radiative transfer equation depend nonlinearly
on themselves) and non-local (the radiation field at a given point
depends on the radiative processes in the entire
atmosphere). Additional challenge rises from the coupling of radiative
transitions, i.e. a change in the rate of one radiative transition has
an effect on all radiative transitions.

In the polarized case the radiative transfer equation takes the following form \citep {Wittmann1974}:

\begin{eqnarray} 
\frac{d}{ds} \left ( \begin{array}{c}  
I \\
Q \\
U \\
V \end{array} \right )  \nonumber \\
=\left ( \begin{array}{c}  
\epsilon_I \\
\epsilon_Q \\
\epsilon_U \\
\epsilon_V \end{array} \right )
- \left ( \begin{array}{cccc}
\eta_I & \eta_Q & \eta_U & \eta_V \\
\eta_Q & \eta_I & \rho_V & -\rho_U \\
\eta_U & -\rho_V & \eta_I & \rho_Q \\
\eta_V & \rho_U & -\rho_Q & \eta_I \end{array} \right)
 \left ( \begin{array}{c}  
I \\
Q \\
U \\
V \end{array} \right ) 
\end{eqnarray}

or in compact notation:

\begin{eqnarray} 
\frac{d \mathbf{I}}{ds} = \mathbf{e} - \mathbf{KI},
\end{eqnarray}
where $\mathbf{e}$ is the emission vector that accounts for the
contributions of spontaneous emission to the intensity and
polarization. The propagation matrix is $\mathbf{K}$. Its diagonal
elements, $\eta_I$, are the absorption terms. The dichroic terms,$\eta_Q$, $\eta_U$ and
$\eta_V$, relate the intensity with the
polarized states, and the dispersion terms, $\rho_Q$, $\rho_U$ and
$\rho_V$, couple the different polarized states. These seven independent
coefficients are functions of the frequency and other parameters
determined by the physical state of the atmosphere, such as the
angles defining the direction of the magnetic field.

We used a nLTE Stokes synthesis and inversion code
\citep {Socas-Navarro+others2000} to synthesize the Ca IR triplet
line profiles for the 2D dynamic atmosphere. The code assumes
statistical equilibrium and complete redistribution. The atmosphere is
dominated by strong UV lines and continua while the lines with large
Zeeman splitting are relatively weak and thus cannot produce
significant changes in level populations. This justifies the use of
the field-free approximation \citep {Rees1969}, i.e. that the
magnetic field has no effect on atomic level populations, instead the
magnetic sublevels are proportionally populated according to their
degeneracy. This reduces the unknowns in the transfer equation to the
same unknowns as in the unpolarized case. Atomic polarization can
have an effect on the emergent Stokes Q and U profiles, especially in
the weak field regime \citep {MansoSainz+TrujilloBueno2003}. Because
of this we have chosen to consider only Stokes I and V in what
follows.

In practice, the code first solves the non-magnetic equation by using
the short characteristics method \citep {Kunasz+Auer1988}. Then the
formal solution is used to compute the emergent Stokes profiles. The
end product is a 2-D time series with the four Stokes parameters for
the Ca IR triplet lines and a photospheric Fe I line at 8497 \AA\
blended with the wing of the 8498 \AA\ line. We do not discuss the Fe
I line, instead we focus on the more dynamic Ca lines. Table
~\ref{tab:lines} shows the atomic level configurations and energies
for the transitions. Shown in the table are also the heights where the
Ca line core optical depths equal unity in the initial simulation
atmosphere as well as the minimum, maximum, and mean heights during the
entire simulation.

\section{Results from the numerical simulation}
\label{results}

Asymmetries in line profiles are caused by atmospheric gradients in
velocity and magnetic field vector. A velocity gradient is always required
for an asymmetry. The line absorption profile at a given height is
centered around a wavelength given by the Doppler shift of the local
fluid velocity. Negative velocities above the main line formation
height shift the absorption profile to the blue, causing more
absorption and an increase in optical depth at the blue wavelengths
while there is less absorption and a smaller optical depth in the red
wing. The opposite is true for positive velocity gradients. The shift
ultimately results in asymmetries visible in Stokes $I$, $Q$, $U$, and
V. Intensity asymmetries due to propagating waves have been discussed
in detail for the Ca H and K lines by Carlsson and Stein (1997). Large
Stokes $V$ area asymmetries are the product of both velocity and
magnetic field vector gradients while a velocity gradient alone can produce
a large amplitude asymmetry \citep {SanchezAlmeida+Lites1992, Landolfi+Landi1996}. In our simulations, because the magnetic field
is weak, the velocity gradients have a larger influence than the
magnetic field gradients on the formation of the asymmetries. As a
result the differences in the shapes of Stokes $V$ profiles along the
horizontal direction are small.

The area asymmetry of a Stokes $V$ profile can be defined as \citep {MartinezPillet+others1997}:

\begin{eqnarray*}
\sigma A=s\frac{\int_{\lambda_0}^{\lambda_1} V(\lambda)d\lambda}{\int_{\lambda_0}^{\lambda_1} |V(\lambda)|d\lambda},
\end{eqnarray*}
where $s$ is the sign of the blue lobe. Because all three lines have a
considerable amount of signal in the wings of the Stokes $V$ profiles,
determining the integration range is not trivial. We do not integrate
the line profile over the full wavelength range, since that can not be
done with real observations. Instead, we chose to determine the wavelength
range from the line intensity profiles by stepping toward the red
starting from the line core until the intensity is equal to the
intensity of the blue wing at 600 m\AA\ from the core. This ensures
that the integration range is centered around the line core, but at
the same time a significant amount of signal in the line wings is
included in the integration range. The signal in the wings, especially
in the 8498 \AA\ line, has usually very little asymmetry and can
overshadow the line core signal making the computed asymmetries
smaller than what they would be if only the line core was considered.

The amplitude asymmetry is given by \citep {MartinezPillet+others1997}:

\begin{eqnarray*}
\sigma a= \frac {a_b-a_r}{a_b+a_r},
\end{eqnarray*}
where $a_b$ and $a_r$ are the unsigned extrema of the blue and red
lobes of the Stokes $V$ profile.

\subsection{Stokes profiles in the initial atmosphere}

\begin{figure*}
\epsscale{1}
\plotone{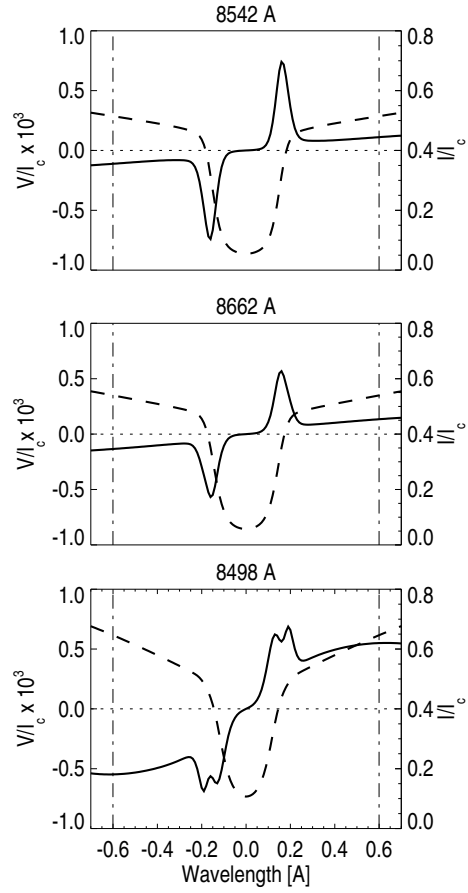}
\caption{Stokes $I$ and $V$ profiles for the Ca lines in the initial atmosphere. In all three panels the left ordinate shows the scale for the solid line (Stokes $V$) and the right ordinate for the dashed line (Stokes $I$). Both Stokes $I$ and $V$ are referred to the quite Sun continuum. The vertical dash-dotted lines show the integration range used in defining the area asymmetry.    
\label{fig:profiles0}
}
\end{figure*}

Since there are no velocity gradients in the initial atmosphere, the
Stokes $I$ and $V$ profiles for the Ca IR triplet lines are symmetric
(figure \ref{fig:profiles0}) in the beginning of the simulation. The
8498 \AA\ Stokes V profile differs from the 8662 and 8542 \AA\
lines. All three have a considerable amount of signal in the wings,
but the Stokes $V$ signal in the 498 \AA\ line wing is almost as strong
as in the lobes of the profile. Furthermore, the lobes of 8498 \AA\
have self-reversal like structures which are not seen in the two other
lines. These structures are caused by the magnetic field gradients in
the atmosphere. The gradients decay with height, so that the 8498 \AA\
line is formed in a region with a stronger gradient than the two other
lines. If the magnetic field were instead constant with height the
8498 \AA\ Stokes $V$ profile would look very much like the 8662 and 8542
\AA\ profiles.

\begin{figure*}
\epsscale{1}
\plotone{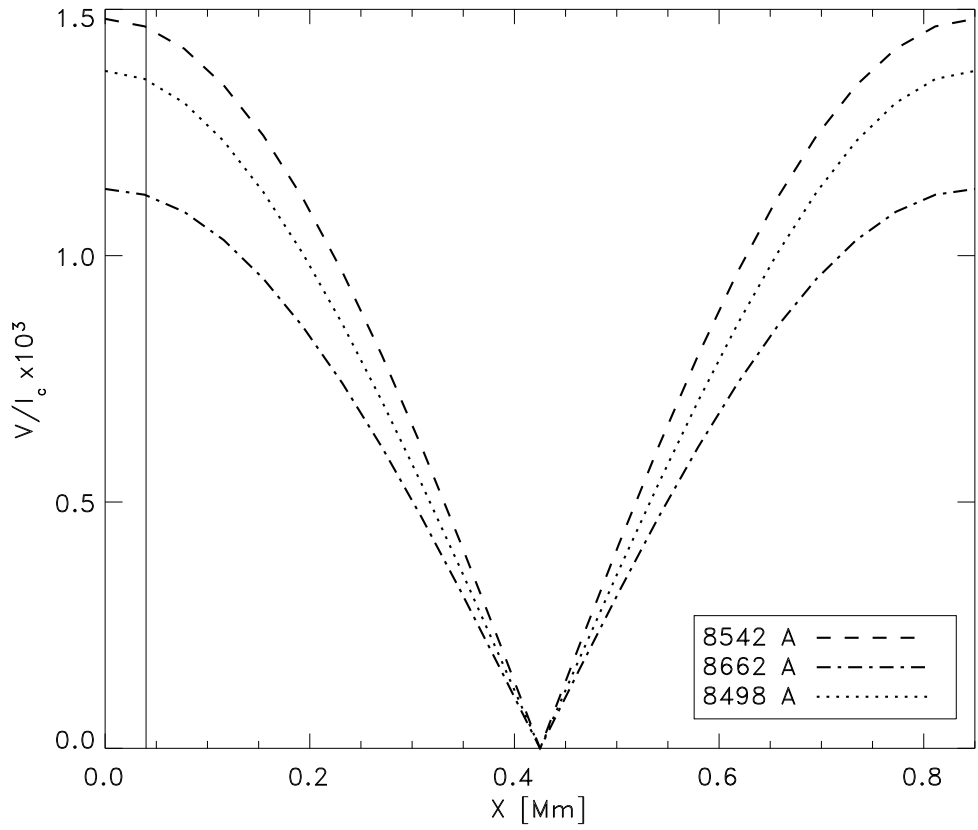}
\caption{Stokes $V$ amplitude as a function of horizontal position in the initial atmosphere. The Stokes $V$ amplitudes are referred to the continuum intensity. The vertical line shows the position where the individual profiles shown in the paper are from.       
\label{fig:ampt0}
}
\end{figure*}

The Stokes $I$ profile amplitude and shape do not change as a function
of horizontal position since we used the field-free approximation. In
a weak magnetic field Stokes $V$ should be roughly proportional to
$\frac{dI}{d\lambda}$, though strict proportionality holds only in the
absence of gradients. The field in the simulation is weak, and the
large magnetic field gradients are in the photosphere, so it is not
surprising that the shape of the Stokes $V$ profiles does not change
much in the horizontal direction. The only visible difference is the
amount of Stokes $V$ signal in the line wings. There is more signal in
the line wings if the line is formed at the edges of the simulation
domain than if it is formed in the middle of the domain. This is true
for all three lines. The $z$-component of the magnetic field changes
sign in the middle and so do the Stokes $V$ profiles as well. The
Stokes V amplitudes in the initial atmosphere as a function of
horizontal position are shown in figure \ref{fig:ampt0}. They are
largest at the sides of the domain where the field is entirely
vertical. The 8498 and 8542 \AA\ lines have roughly the same
amplitudes even though they are formed at different heights. The
effective Land\`e $g$-factors for the 8498, 8662, and 8542 \AA\ lines
are 1.07, 1.10, and 0.83 respectively, so it is not surprising that
the 8542 \AA\ Stokes $V$ signal has almost the same amplitude as the
8498 \AA.

\subsection{Time evolution of the Stokes profiles}

\begin{figure*}
\epsscale{1} 
\plotone{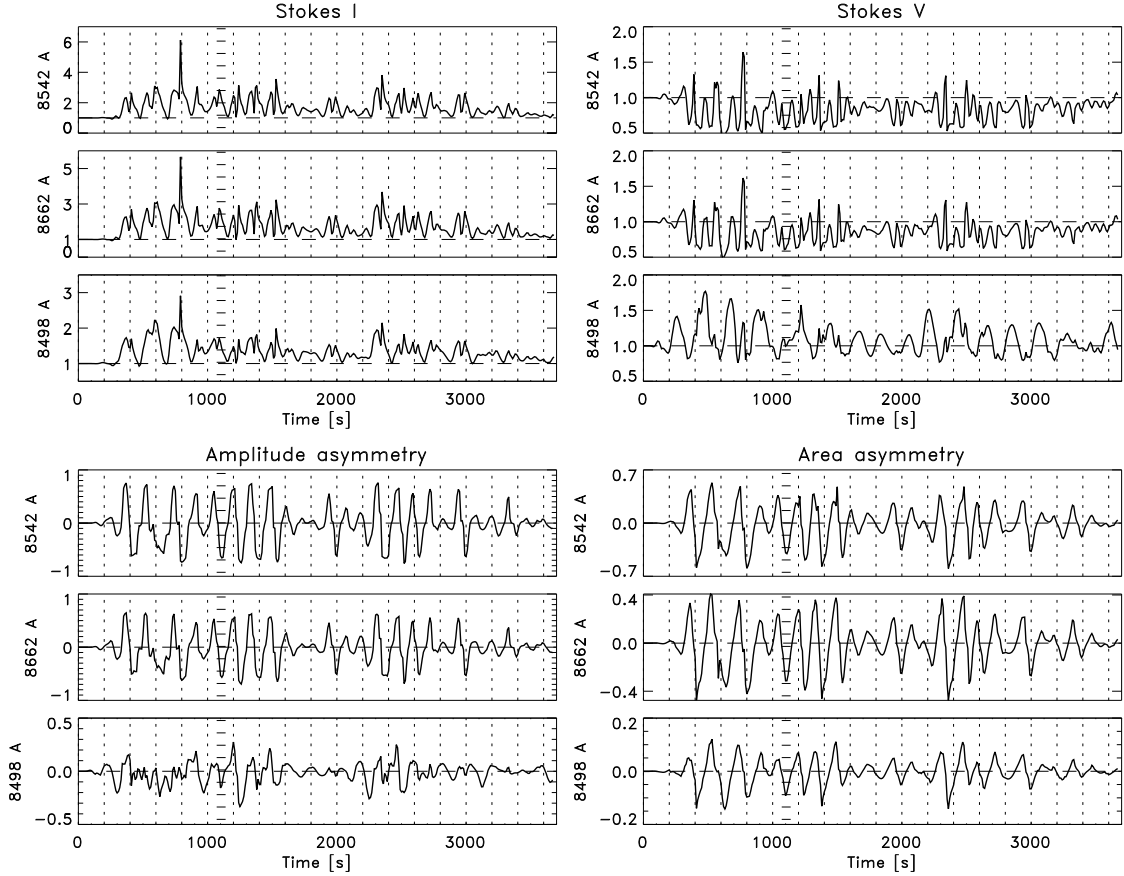}
\caption{Time evolution of Stokes $I/I(\textrm{t=0})$ (measured at line core) and Stokes $V$/$V$(\textrm{t=0}) amplitude, Stokes $V$ area and amplitude asymmetries at the left
edge of the simulation domain. The vertical dashed lines are located
at every 20 seconds interval. The region of horizontal lines shows the
location of the shock discussed in detail in the text.
\label{fig:amp&asym}
}
\end{figure*}

Waves propagating through the formation regions of the lines cause
oscillations in Stokes $I$ and $V$ amplitude as well as Stokes $V$
amplitude and area asymmetries (figure \ref{fig:amp&asym}). The Stokes
$I$ oscillation time series are quite similar in all three lines
except for the slightly more saw tooth-like and less smooth appearance
in 8662 and 8542 \AA\ line time series. The large intensity peak
visible in all three lines at about 800 seconds coincides with the
smallest plasma-$\beta$ value in the simulations thus coinciding with
a shock causing a larger than normal compression of plasma. In general
an increase in intensity, i.e. less absorption, is associated with a
decrease in Stokes $V$ amplitude. This relationship is most clearly
seen in the 8498 \AA\ time series. Small variations barely visible in
the 8498 \AA\ Stokes $V$ amplitude time series are prominent in the
8662 and 8542 \AA\ lines. The overall shape of the asymmetry time
series resembles the shape of the piston (figure \ref{fig:piston}),
i.e. fluid velocity as a function of time. The growing amplitude of
the propagating waves is manifested in the increased amplitude of the
asymmetry oscillations: the asymmetry oscillation of the 8498 \AA\
line has a much smaller amplitude because it is formed lower.

\begin{figure*}
\epsscale{1}
\plotone{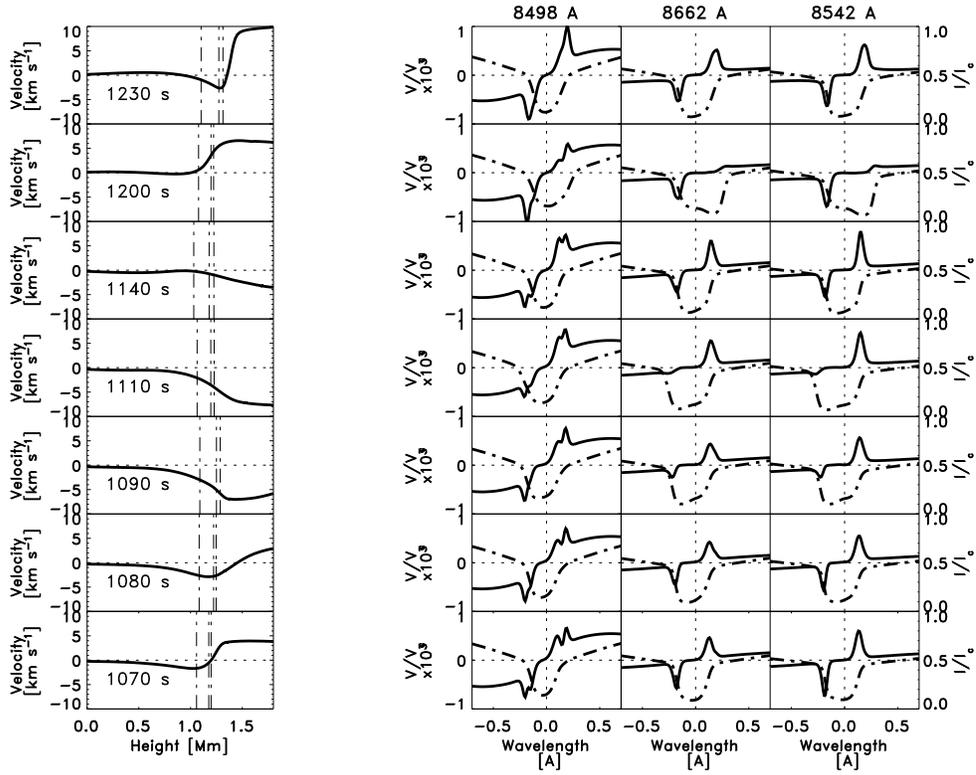}
\caption{Time evolution of a shock. Each row in the figure represents a snapshot of the simulation. The time for the snapshot is shown in the upper part of the first panel of each row. Shown are the velocity in the atmosphere as a function of height and the Stokes $V$ (left ordinate, solid line) and Stokes $I$ (right ordinate, dash-dotted line) profiles for the three Ca lines. The dashed lines in the velocity panels show the height where the Ca line center optical depths equal unity.  
\label{fig:shock}
}
\end{figure*}

Figure ~\ref{fig:shock} shows the time evolution of the atmospheric
velocity and Ca lines as a steepening wave propagates through the
atmosphere. At first (time 1070 s) the velocity in the region where
the Ca lines are mostly formed is negative and has a small amplitude
resulting in an absorption profile that is only slightly
blue-shifted. The shift is visible in the intensity profiles, but not
in Stokes $V$. This is because the shift is not yet large enough to
affect the blue lobe in Stokes $V$. The 8498 \AA\ line is affected by
the upward propagating wave before the other two Ca lines (time 1080
s). The velocity in the Ca 8498 \AA\ formation region grows and the
absorption profile shifts even more to the blue so that now not only
the Stokes $I$ but also $V$ is more asymmetric ($I$ more than
$V$). Later in time (time 1090 s) the 8662 and 8542 \AA\ intensity
profiles also start to become more asymmetric as the wave propagates
upward. Finally (time 1110 s), the velocity gradients are large enough
to cause the blue lobes of the Stokes $V$ profile in 8662 and 8542
\AA\ disappear completely. The 8498 \AA\ line is formed lower where
the gradients are not yet large enough to cause the lobe to
disappear. Before the next wave arrives, the velocity amplitude
becomes very small and the blue lobes start to gradually reappear
until the Stokes $V$ profiles are again fairly symmetric (time 1140
s). The next wave brings a positive velocity that causes the
absorption profile to shift toward red and eventually the red lobes of
the 8662 and 8542 \AA\ Stokes V signals disappear (time 1200 s). The
lobes reappear once the velocity becomes again very small and the
absorption profile is centered around the non-shifted line core (time
1230 s).

This pattern of disappearing and reappearing Stokes $V$ lobes repeats
itself throughout the simulation. When the atmosphere is close to
rest, an amplitude peak associated with a symmetric profile occurs. 

The amplitudes of the Stokes $V$ asymmetries, both area and amplitude,
are closely tied to the amplitudes of the velocity gradients present
in the atmosphere. In figure ~\ref{fig:velgrad} are plotted the
asymmetries as functions of the atmospheric velocity gradient. The
velocity gradient is defined as the difference between the velocity at
two heights, 0.5 and 1.3 Mm. A negative gradient indicates a velocity
that increases upward and vice versa. The correlation between the
asymmetries and the velocity gradients is stronger in the 8662 and
8542 \AA\ lines. If the velocity gradient was determined lower in the
atmosphere there would be a stronger correlation in the
case of the 8498 \AA\ line. The way the velocity gradient is defined
works well most of the time. But there are cases when the definition
fails and gives rise to the outlier points in figure
~\ref{fig:velgrad}, where a small asymmetry is associated with a very
large velocity gradient. These large velocity gradients are actually
very localized and at a height of 1 Mm and above, i.e. above the
line's main formation region. The line in these cases is
formed mostly in a region where the velocity profile is nearly flat
and does not cause large asymmetries in the line. Area and amplitude
asymmetries may prove to be good diagnostics for atmospheric
velocity gradients.

\begin{figure*}
\epsscale{1}
\plotone{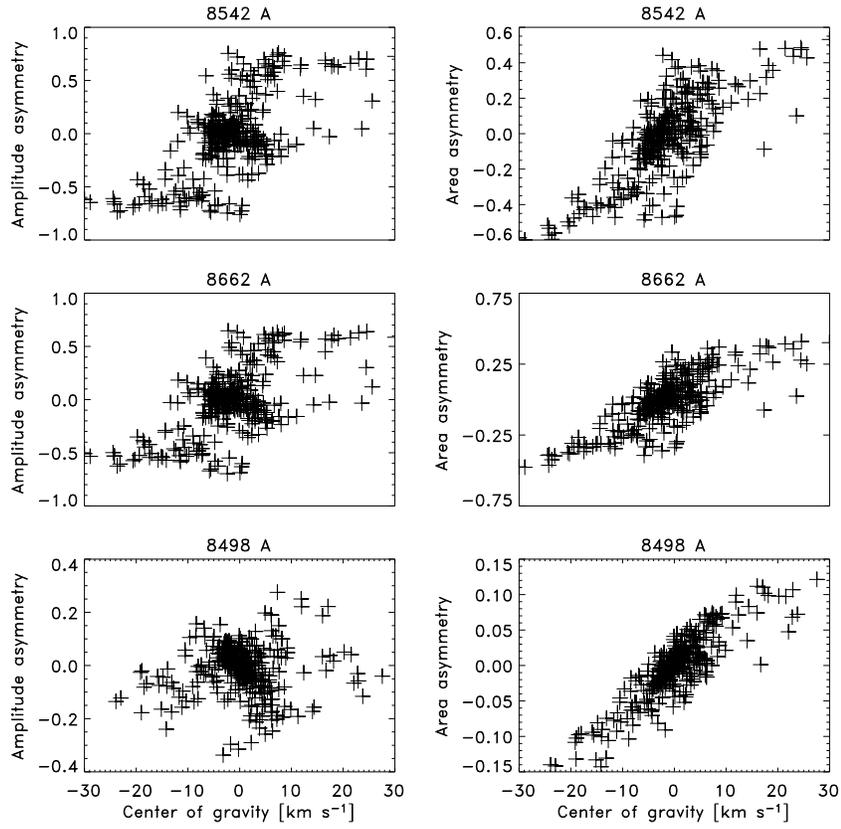}
\caption{Amplitude and area asymmetries as a function of the center of gravity for the Ca lines.    
\label{fig:cog}
}
\end{figure*}

Not surprisingly, there is a similar relationship between the
area asymmetries and the center of gravities of the Stokes $I$ profiles
(fig. \ref{fig:cog}). There is much more scatter in the case of amplitude asymmetries and line center of gravities. The center of gravity for a profile is defined
as:

\begin{eqnarray*}
cog=\frac{\int_{\lambda_0}^{\lambda_1} (I_c-I)\lambda d\lambda}{\int_{\lambda_0}^{\lambda_1} (I_c-I)d\lambda},
\end{eqnarray*}
where $I_c$ and $I$ are the continuum and line intensities,
respectively. The center of gravity is actually related to an average of the bulk velocity. In this simulation, the times of maximum velocity gradient are also the maximum excursions of the bulk velocity, which explains the good correlation between the center of gravity and profile asymmetry. 

\begin{figure*}
\epsscale{1}
\plotone{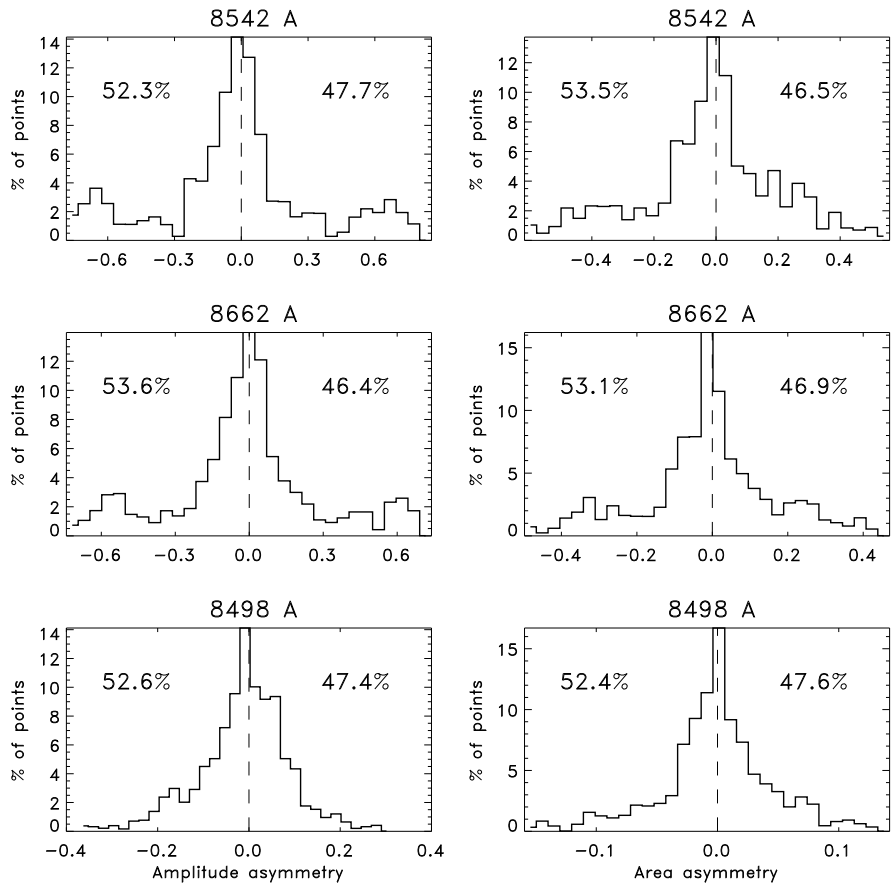}
\caption{Histograms of amplitude and area asymmetries in the Ca lines. The percentages show the amount of negative and positive asymmetries.   
\label{fig:histogram}
}
\end{figure*}

Histograms of the amplitude and area asymmetries (figure
\ref{fig:histogram}) are fairly symmetric. The three Ca lines have slightly more negative
than positive asymmetries i.e. the Stokes $V$ blue lobe has in
general a larger amplitudes than the red lobe. In the amplitude asymmetry histograms of the 8662 and 8542
\AA\ lines peculiar side lobes can be found. These lobes are not seen
in the 8498 \AA\ line where the histograms decay steadily on both
sides. The lobes correspond to the extremely asymmetric Stokes V
profiles associated with shocks. The transition from a symmetric
Stokes $V$ profile to a strongly asymmetric profile takes less time
than the symmetric and strongly asymmetric phases. This is seen in
the histograms where there are fewer moderately
asymmetric profiles than symmetric and strongly asymmetric
profiles. The cores of the area asymmetry histograms (right-most panels of figure \ref{fig:histogram}) for the three Ca lines are similar. However, the
8662 and 8542 \AA\ histograms have wings that extend to approximately
-0.4 and 0.4 whereas the 8498 \AA\ area asymmetries are considerably
smaller. The extended wings are caused by the strongly asymmetric
profiles associated with shocks.

\subsection{Discussion}

Waves propagating through and forming shocks in the formation regions
of the Ca lines cause a time-varying pattern in the Stokes $V$
profiles. As the local fluid velocity changes with the changing phase
of the propagating wave, the line absorption profile is shifted away
from the line core. This combined with velocity gradients in the
atmosphere leads to a repeated pattern of disappearing and reappearing
lobes in the Stokes $V$ profiles. In the simulation large asymmetries
in the Stokes $V$ profiles are commonly found. Since the 8662 and 8542
\AA\ lines are formed higher up in the atmosphere, where the wave
amplitudes are larger, the time varying pattern is stronger in them
than in the 8498 \AA\ line. Similar dynamic behavior of Stokes $V$
profiles as in our simulation has been seen in a 2D MHD simulation of the
photosphere by \citet{Steiner+others1998}. They simulated photospheric
convection and its interplay with flux sheets in the intergranular
lanes. The effect of the swaying flux sheet and passing shocks was
visible in both the shape and wavelength of the Stokes $V$ profiles.

\begin{figure*}
\epsscale{1}
\plotone{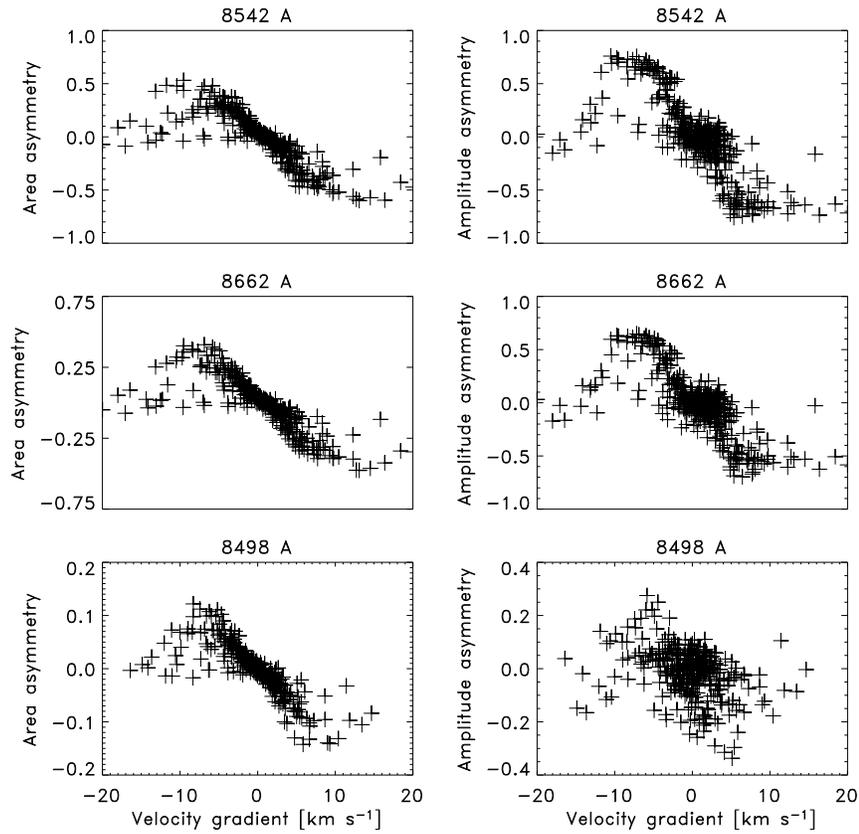}
\caption{Amplitude and area asymmetries as a function of the velocity gradient. The gradient is determined in the upper photosphere/chromosphere, 0.5-1.3 Mm.   
\label{fig:velgrad}
}
\end{figure*}

Earlier work \citep[e.g.][] {SanchezAlmeida+others2003, Ploner+others2001,Grossmann-Doerth+others2000, Steiner+others1998, Grossmann-Doerth+others1989, Grossmann-Doerth+others1988} addressing the formation of
large asymmetries in Stokes $V$ profiles in the photosphere found that
variations along the line-of-sight and polarity reversals in the
atmosphere are important in producing high asymmetries. In our
simulation the photospheric iron line in the wing of the 8498 \AA\ Ca
line has very little asymmetry in it. Since the piston used in the
simulation is homogeneous over the domain, asymmetries can not rise
from line profiles being shifted relative to one another. Only the
magnetic field and velocity gradients are able to produce asymmetries,
but the wave amplitude in the photosphere is not large enough to make
the iron line highly asymmetric.  This is in accordance with previous
work where gradients in velocity and magnetic field alone seldom
produce large asymmetries in photospheric lines. The situation is
somewhat different for the chromospheric lines. Highly asymmetric and
even single-lobed Stokes $V$ profiles are a fairly common occurrence
in the simulated Ca lines even though no polarity reversals take place
in a single ``pixel'' and the line-of-sight is along the vertical
axis. Unlike in the photosphere, a combination of a velocity gradient
and a magnetic field gradient is enough to produce single-lobed Stokes
$V$ signals in the chromosphere. The strong dependence of the area and
amplitude asymmetries on the velocity gradients shown in figure
\ref{fig:velgrad} was heuristically explained for the LTE case by
\citet{Grossmann-Doerth+others1989}: the relation between the total
change of Zeeman shift and line Doppler shift in the lines formation
region governs the magnitude of the profile asymmetry. This relation
has been seen in observations as well: for example,
\citet{Balasubramaniam+others1997} found a clear correlation in the
6301 and 6302 \AA\ iron line center-of-gravities and Stokes $V$
amplitude asymmetries.

The strong similarity in the Stokes $V$ profile shapes at a given time
along the horizontal direction shows that in the high-$\beta$ regime
the effect of chromospheric magnetic field gradients on the
asymmetries is not as strong as the effect of the velocity
gradients. If the magnetic field scale height is set to be 4.25 Mm
instead of 0.85 Mm, so that the magnetic field gradient is larger
higher up, there still is very little difference in profile shape
along the x-axis. The velocity gradients clearly dominate the
formation of the asymmetries in the simulation.

 If the Ca lines are formed in a region where acoustic waves are
important and the field is only moderately strong we would expect to
see this pattern in observations as well. Stokes $V$ asymmetries may
prove to be a possible diagnostic for estimating velocity gradients in
the lines' formation regions for fairly quiet-Sun areas with moderate
magnetic flux. Spatial resolution will be a serious issue in trying
to observe dynamic behavior similar to the simulation. Because of
limited spatial resolution, polarity reversals and magnetic field
distributions within a pixel will easily smudge out the asymmetries in
the Stokes $V$ profiles and decrease the signal amplitude. The
comparison of observations with the simulated profiles is therefore
not likely to be straight-forward. It should still be possible to
observe wave propagation in the Stokes $V$ profiles of the Ca IR
triplet lines though one should not necessarily expect to see
asymmetries as extreme as in the simulations.

The next step is to test if this simple simulation is able to
reproduce line profiles observed in the quiet Sun. In the future, we
plan to use quiet Sun observations of the Ca IR triplet lines from the
SPINOR instrument \citep{Socas-Navarro+others2005}. Another
interesting question to pursue is how a stronger magnetic
field would affect the atmospheric dynamics and consequently the line
profiles and their time-dependent behavior. Since a 3D MHD code with
full radiative transfer is not available, using a similar approach as
here, i.e. combining a 2D MHD code with radiative cooling and the
Stokes synthesis code, is an attractive alternative.

Understanding the formation of the Ca IR triplet lines in a dynamic
atmosphere will hopefully lead to a better interpretation of
observations and ultimately give us new insight into the dynamics of
the magnetic quiet Sun.

\acknowledgments We are greatful to J.M. Borrero for constructive comments on the manuscript.

\begin{deluxetable}{lllllllll}
\tablewidth{0pt}

\tablecaption{Spectral lines.}

\tablehead{ &Level  & $E_i$ & $E_k$ & log$gf$ & Z(0) & min(Z) & max(Z) & mean(Z)\\
 &Configuration  & [cm$^{-1}$] & [cm$^{-1}$] & & [Mm] & [Mm] & [Mm]     & [Mm]}
\startdata
\ Ca II 8542 \AA  & $^2P^O_{3/2}$-$^2D_{5/2}$ & 13711 & 25414  & -0.36  & 1.24           & 0.89   & 1.43 &1.19       \\
\ Ca II 8662 \AA  & $^2P^O_{1/2}$-$^2D_{3/2}$ & 13650 & 25192  & -0.62  & 1.16  	  & 0.87 & 1.36 & 1.15       \\
\ Ca II 8498 \AA  & $^2P^O_{3/2}$-$^2D_{3/2}$ & 13650 & 25414  & -1.32 & 0.97  	  & 0.80 & 1.17 & 1.02       \\

\ Fe I  8497 \AA   & $^3F_{3}$-$^3F_2$ & 37024 & 48797 & -0.95      &
  &      &      &  \\
\enddata
\label{tab:lines}
\end{deluxetable}

\end{document}